\documentclass[%
  prb,%
  letterpaper,%
  twocolumn,%
  showpacs,%
  floatfix,%
  superscriptaddress%
]{revtex4-1}

\RequirePackage{graphicx,amsmath,amssymb,bm}
\RequirePackage{hyperref,verbatim}
\hypersetup{%
  breaklinks = {true},
  citecolor = {blue},
  colorlinks = {true},
  linkcolor = {red},
  pdfcreator = {\LaTeX\ and \flqq hyperref\frqq},
  pdffitwindow = {true},
  pdfmenubar = {true},
  pdfpagelayout = {SinglePage},
  pdfstartview = {Fit},
  pdftoolbar = {true},
  plainpages = {false},
}

\usepackage{siunitx}

\newcommand{\Fref}[1]{Fig.~\ref{#1}}

\renewcommand{\eqref}[1]{Eq.~(\ref{#1})}

\RequirePackage[usenames,dvipsnames]{xcolor}
\usepackage{soul}

\begin{document}

\title{Zigzag phosphorene nanoribbons: one dimensional resonant
channels in two dimensional atomic crystals}

\author{C. J. P\'aez}
\affiliation{Faculdade de Ci\^{e}ncias Aplicadas, Universidade
Estadual de Campinas, 13484-350 Limeira, SP Brazil}%
\author{D. A.  Bahamon}%
\email[Corresponding author:
]{dario.bahamon@mackenzie.br}
\affiliation{MackGraphe -Graphene and Nano-Materials Research Center, Mackenzie
Presbyterian University,
Rua da Consola\c{c}\~{a}o 896, 01302-907, S\~{a}o Paulo, SP, Brazil}%
\author{Ana L. C. Pereira}%
\affiliation{Faculdade de Ci\^{e}ncias Aplicadas, Universidade
Estadual de Campinas, 13484-350 Limeira, SP Brazil}%
\author{P. A. Schulz}%
\affiliation{Faculdade de Ci\^{e}ncias Aplicadas, Universidade
Estadual de Campinas, 13484-350 Limeira, SP Brazil}%

\begin{abstract}
We theoretically investigate phosphorene zigzag nanoribbons as a platform for constriction engineering. In the presence of a constriction at one of the edges, quantum confinement of edge protected 
states reveals conductance peaks, if the edge is uncoupled to the other. If the constriction is narrow enough to promote coupling between edges, it gives rise to Fano-like as well as anti-resonances 
in the transmission spectrum. These effects are shown to mimic an atomic chain like behavior in a two dimensional atomic crystal.
\end{abstract}


\maketitle

\section{Introduction}

Low dimensional systems have attracted the attention at least in the past fifty
years, since the development of semiconductor epitaxial growth and deposition of
metallic thin films \cite{5391729}. The early scenario, back in the 1960's, as promising as it
appeared, has evolved into a mainstream interest in condensed matter physics due
to landmark discoveries in the late 1970s and early 1980s, like the quantum Hall
effect \cite{Klitzing:1980qy} and the conductive polymers \cite{C39770000578},
respectively 2D and 1D systems. The
subsequent discovery of new carbon allotropes, showing stable structures either
in 0D (fullerenes), 1D (carbon nanotubes) and 2D (graphene) consolidated this
scenario in an exciting research field \cite{carbonall}. The isolation of strictly one atom thick
layers in the first years of the present century opened a wider window for both
basic physics and device applications \cite{graphroad}. These new disruptive research efforts,
initially impulsed by graphene, are
nowadays detaching from carbon based roadmaps, as also envisaged, at the
beginning of the graphene boost, by Novoselov, Geim and coworkers \cite{NovoselovPNAS}.

A very recent 2D atomic crystal  of black phosphorous \cite{Li:2014qf,Koenig,nn501226z,nl5008085}, namely
phosphorene, is a promising system in
which 2D properties together with strictly 1D chain behavior are present in
different energy windows, hence allowing to a same device to be tuned from a 1D
to a 2D system by simply tuning the Fermi energy.
In the present work we focus on the 1D energy window, created by effective doubly degenerate band (in the relevant energy scale)
 associated to states strongly localized at the zigzag
edges \cite{Ezawa,Carvalho} of phosphorene nanoribbons, whose properties are explored using a new strictly one
dimensional resonant tunnelling device. 

The double barrier resonant tunnelling device \cite{5391729,1073171,1655067},
conceived here as an atomically
precise segmentation at one of the edges, shows unusual geometry, since the
direction of the barriers is perpendicular to the well and contact regions\cite{102079,354377}.
Among our
findings we show that for a thin barrier case (constriction with narrow step from  the upper zigzag edge), the resonant tunnelling permits a
spectroscopy of the band structure of phosphorene nanoribbons in this energy
window. Furthermore, progressive widening of the barriers (enhancing the step width of the constriction) therefore nearing the constriction to the other edge leads to edge coupling
effects featuring resonances with Fano line shapes \cite{FanoResRMP,HuaLY15,Crespi:2015kq}
revealing also a new discrete/continuum states coupling system. For this latter coupled edge system, the
transmission probability characteristics turn out to present clear features of
both (i) the actual finite confining segment coupled to an infinite (not
segmented) edge and, (ii) the properties of an infinite narrow nanoribbon with strongly coupled
edges. These results are independent of the area of the device region, solely
from the segmented region length and distance between the edges, revealing an
effective chain-like behavior of the nanoribbon's edges.

In what follows, we initially discuss the ``bulk" electronic properties of a phosphorene nanoribbon, presenting the model calculation framework, as well as the effects of edge coupling on the 
conductance of these
infinite zigzag ribbons that are essential to understand the resonant tunnelling
spectra. Next, the geometry of the actual investigated segmented device is presented, introducing the resonant tunnelling effects.
In the sequence, the core of the results is devoted to
explore the segmented edge device resonant tunnelling behavior, showing how the defects on the edge may actually enrich the scenario instead of solely washing out the announced effects, giving 
further 
evidences that the phenomenon is restricted to the  atoms at the very edge. Finally, the conclusions suggest a bridge between the present 1D systems embedded in a 2D crystal and ongoing research on 
isolated atomic chains.

\section{Phosphorene zigzag nanorribons and model calculation: edge coupling effects}

\begin{figure} [h]
\includegraphics[width=1\columnwidth]{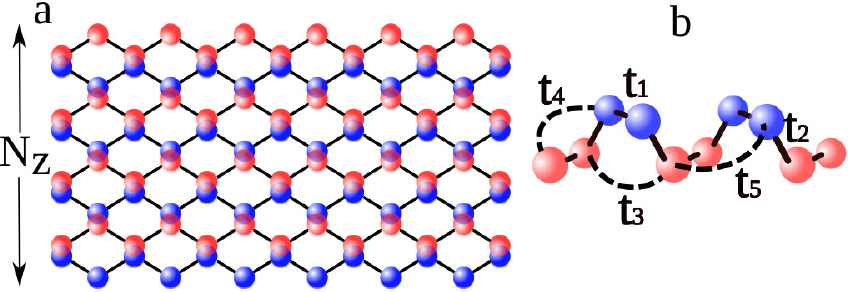}
\includegraphics[width=1\columnwidth]{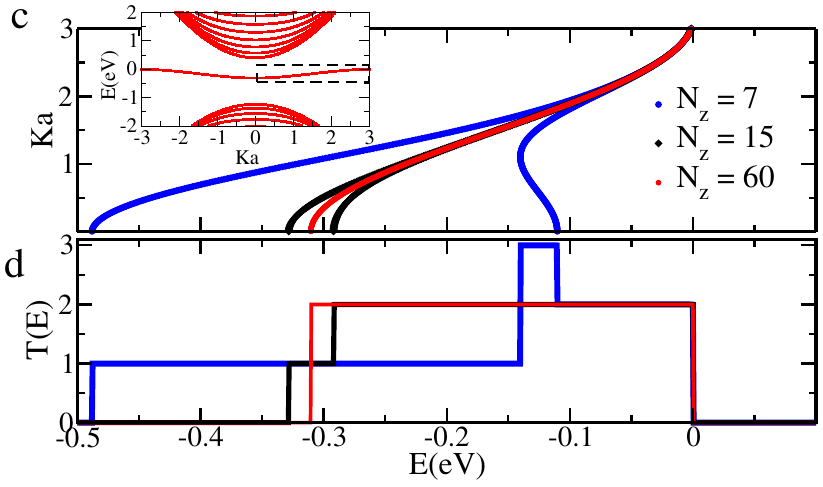}
\includegraphics[width=1\columnwidth]{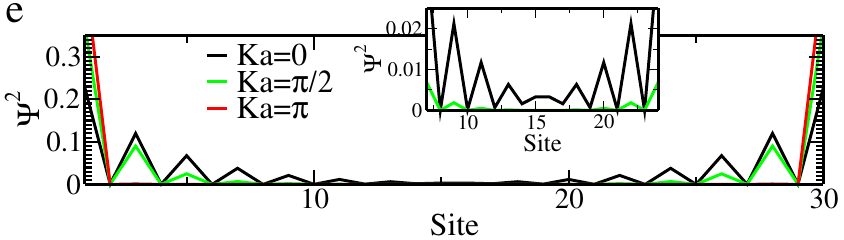}
\caption{ (Color online) \textbf{(a)} Illustration of a zigzag phosphorene
nanoribbon
of width $N_z$, where $N_z$ is the number zigzag chains. \textbf{(b)} Hopping
parameters  used in the four band model \cite{Rudenko:2014by}.
\textbf{(c)} Structure of the central band of edge states of nanoribbons with different widths. 
The inset includes the band structure of the bottom and top of the conduction and valence 
bands for the $N_z = 60$ case. \textbf{(d)} Transmission probabilities for the edge states depicted in \textbf{(c)}. \textbf{(e)} 
Probability amplitude of edge states of the $N_z = 15$ case for different ka values. The inset 
reveals the stronger non-zero overlapping of states of the two edges at the bottom of the 
central band, $ka =0$.}
\label{fig:Fig1}
\end{figure}

The essential atomistic aspects of the structures investigated are depicted in
\Fref{fig:Fig1}. \Fref{fig:Fig1}(a) shows a segment of an infinite zigzag edged
phosphorene nanoribbon of width $N_Z$ = 8, the number of zigzag chains
along the ribbon in this case. The tight binding hopping parameters
considered, as discussed below, are indicated in \Fref{fig:Fig1}(b).

The quite complex electronic structure of phosphorene, already at energy ranges rather
close to the Fermi energy, hinders a wider use of single orbital tight-binding
models in chasing the alluded electronic and transport properties of systems
based on this new material. Nevertheless, the use of such model is well
validated, by means of comparisons with first principle electronic structure
calculations \cite{Rudenko:2014by}, for the very energy window of interest around the gap, namely a
double central band. This central band for zigzag phosphorene nanorribons has been predicted for phosphorene \cite{Carvalho,Ezawa} and is absent in Graphene.
We use here the same tight-binding parametrization for
phosphorene proposed by Rudenko\cite{Rudenko:2014by} considering a Hamiltonian $H =
\sum_{ij}t_{ij}c^{\dagger}_ic_j$, where $c_i$ ($c^{\dagger}_i$)
is the
creation (annihilation) electronic operator at site $i$ and $t_{ij}$ is
the hopping integral between sites $i$ and $j$.
In this model, to characterize the low energy
electronic  properties five hopping integrals are required \cite{Rudenko:2014by}: $t_1 = -1.220$ eV,
$t_2$ = 3.665 eV, $t_3$ = -0.205 eV, $t_4$ =-0.105 eV and $t_5$ = -0.055 eV .
The transmission $T=\text{Tr}[\Gamma_LG^r\Gamma_RG^a]$ is
calculated
using the recursive Green's function \cite{LewM13} $G^r=[E+i\eta-H-\Sigma_L-\Sigma_R]^{-1}$
in the phosphorene lattice representation. The left and right
contacts broadening function
$\Gamma_{L(R)}=i[\Sigma_{L(R)}-\Sigma_{L(R)}^{\dagger}]$ and the self-energy of
contact $ \Sigma_{L(R)}$
 are calculated recursively for the semi-infinite zigzag phosphorene
nanoribbons~\cite{LopezSancho};
 other electronic properties such as the local density of states (LDOS)
 $\rho_{ii}=-\text{Im}[G^r(\vec{r_i},\vec{r_i},E)]/\pi $ are also calculated.

The electronic and transport properties of a host zizgzag nanoribbon in which a
finite segment will be latter tailored in are also summarized in \Fref{fig:Fig1}.
The inset in \Fref{fig:Fig1}(c) depicts the energy window of interest, showing
the top(bottom) of the valence(conduction) band and an effectively degenerate central
band \cite{Ezawa}. These central bands present cosine like dispersions
characteristic for 1D systems \cite{Ezawa}. Indeed, the degeneracy comes from the fact that
the width of the ribbon here is $N_Z$ = 60, which guaranties that the two edges are  effectively
uncoupled\cite{Sisakht:2015wd}. Hence this width will be chosen for the host ribbon where the constriction will be introduced.

The effect of edges coupling on the band structure can be followed in the main part
of \Fref{fig:Fig1}(c), showing a zoom of the central band energy range. Having
in mind the uncoupled limit of $N_Z$ = 60 (red curve), lifting of the central
band effective degeneracy starts (in the present relevant energy scale) for $N_Z$ = 15 (black curve) at the center of the
Brillouin zone with an approximately symmetric splitting of slightly deformed
cosine-like bands. Indeed an incipient overlap of wave function in this
situation is illustrated in \Fref{fig:Fig1}(e), with a noticeable amplitude of
the wave function in the atomic sites well inside the ribbon (It should be
noticed that $N_Z$ corresponds to the number of zigzag chains, hence for $N_Z$
=15, there will be 30 atomic sites in the unit cell). For
extremely thin ribbons, $N_Z$ = 7, the
splitting becomes of the order of the uncoupled band widths (blue curve). More
striking is the drastic change in the shape of one of the bands, showing a
local
maximum at the center at $Ka$ = 0 and a minimum at $Ka$ = 1.

The consequences of the edge coupling on the transmission probabilities along
the edges are qualitatively significant as can be seen coming back to \Fref{fig:Fig1}(d).
Degenerate bands sum up to a plateau of $T$ = 2 (red curve). A slight lifting
of the degeneracy breaks the lower threshold of the plateau introducing a $T$ =
1 step, an energy range where there is only one conduction channel\cite{Ezawa}. However, extreme
coupling leads to a $T$ = 3 plateau for $-1<ka<1$, where the different states corresponding to the same
energy in this band are added to the channel associated to the other cosine-like band.

\begin{figure} [t]
\includegraphics[width=0.9\columnwidth]{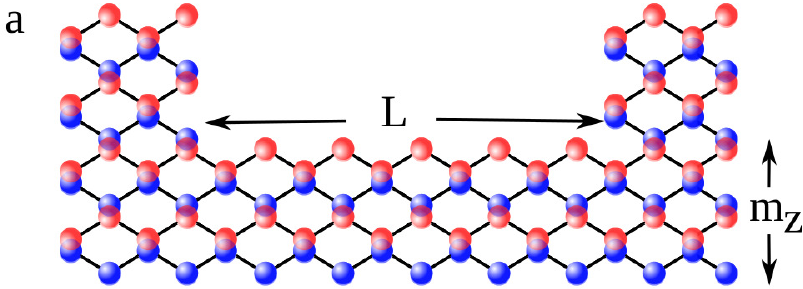}
\includegraphics[width=1\columnwidth]{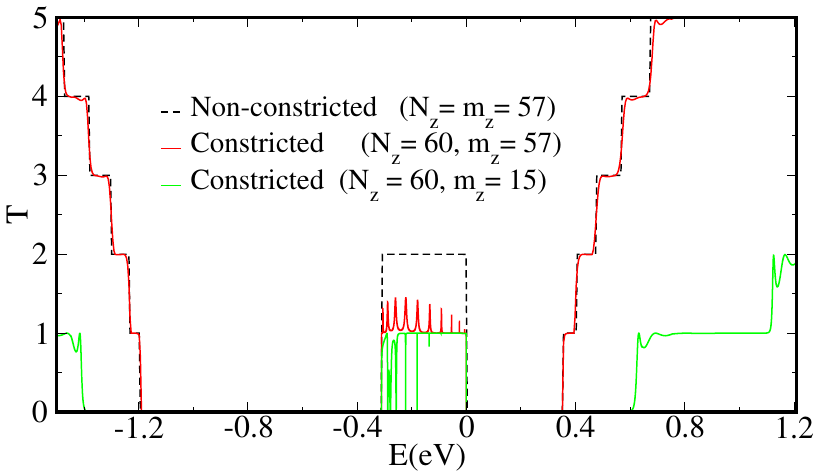}
\caption{
(Color online)\textbf{(a)} Schematic picture of a constriction characterized by the parameters $L$
=10, actually used throughout the work, and $m_z$ = 4, here only for the sake of illustration. \textbf{(b)} 
Transmission probabilities, as a function of the Fermi energy, at the energy range 
corresponding to the central band, including the bottom and top of the conduction and 
valence bands. Three different situations are depicted: two constriction defined at 
nanoribbons of width $N_z$ =60, both with $L$=10, and $m_z$ = 57 (red) and $m_z$ = 15 (green) and a 
zigzag nanoribbon without any constriction, $N_z$ = $m_z$ = 57 (black line).}
\label{fig:Fig2}
\end{figure}

\section{Segmented nanorribons: resonant tunnelling in 1d effective chains structures}
In the energy energy range of the edge states band, the ``bulk" of the nanoribbon acts mainly as ``in plane" 
substrate for the two dimensional channels at the edges. This condition,
evidenced by the electronic band structure discussed in the previous section 
raises the question of a mean to observe experimentally those effective
one-dimensional chains embedded in the rather complex phosphorene crystalline
structure.
In order to test our hypothesis we propose the  segmented nanoribbon structure (constriction) illustrated in \Fref{fig:Fig2}(a). The
segment of a thinner region of the nanoribbon of width $m_Z$, also given in number of zigzag
chains,  is defined by a length $L$ in units of
atoms removed along one zigzag direction. One essential parameter is
the step width between the semi infinite upper edges and the central segment,
which is simply defined as $N_Z - m_Z$ and, as will be seen below, defines the
barrier thickness in the resonant tunnelling.

shows the transmission probabilities through two constrictions of lenght $L=10$ with $N_Z - m_Z = 3$ and $N_Z - m_Z = 45$ step widths, compared to a bare $N_z=57$ 
nanorribon, as a 
function of energy.
In order to avoid any coupling between the edge states the width of the nanoribbon in the contacts is also $N_z = 60$. Transmission
plateaus above (below) the edge states band are 
shown, for the sake of completeness, since these structures are of entirely different character than
the resonances in the central band we will be focusing on. These transmission
plateaus are due to the lateral confinement in a nanoribbon. This is confirmed by the green curve for
$m_z = 15$, a deep step leading to a large shifting of the valence and conduction bands transmission plateaus,
evidencing also the well known Fabry-Perot oscillations\cite{Szafer:1989qe} due to the geometrical modulation; these effects are already well known for graphene  and square lattice nanorribons with 
constrictions \cite{Wakabayashi:2000cs,Wakabayashi:2001hb,Munoz-Rojas:2006tg,YanRL15, Szafer:1989qe}. 

\begin{figure} [h!]
\includegraphics[width=1\columnwidth]{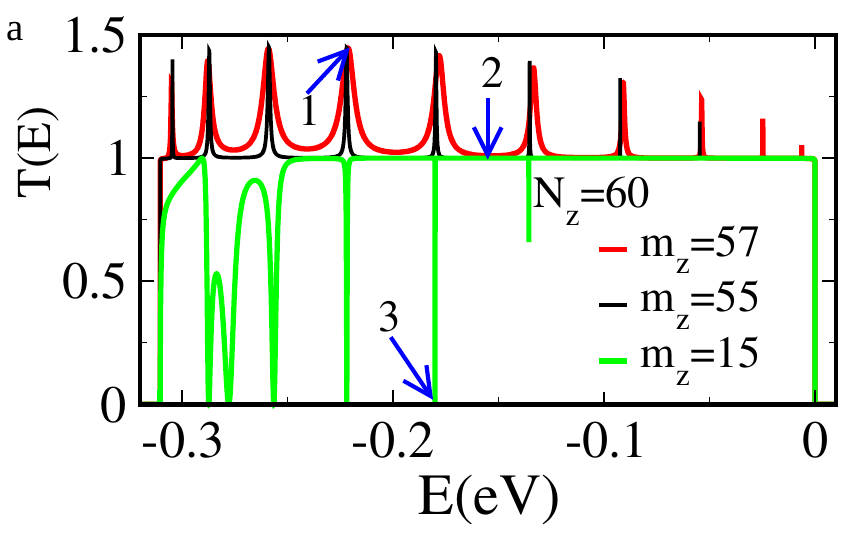}
\includegraphics[width=1\columnwidth]{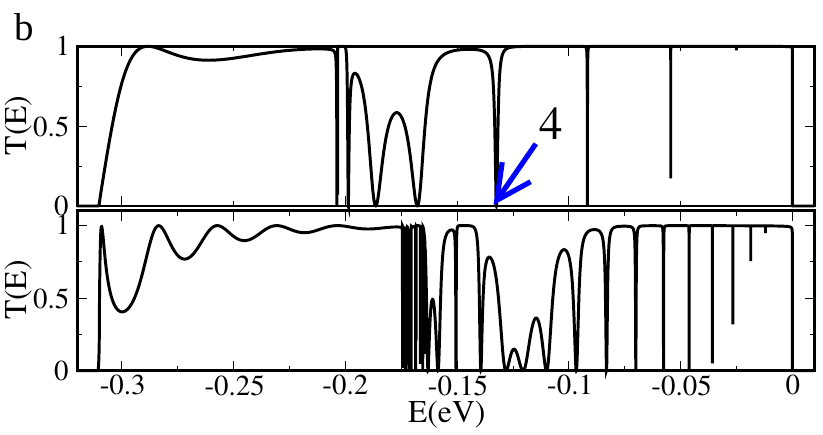}
\caption{(Color online) \textbf{(a)} Transmission probabilities, as a function of the Fermi energy, at the
energy range of the central band corresponding to nanoribbons with constrictions: red and 
green correspond to the cases in \Fref{fig:Fig2}(b): $N_z$ =60, all with $L$ =10 and $m_z$ = 57 ( shallow 
constriction) and $m_z$ = 15 (deep constriction), respectively. The black curve corresponds to a 
constriction with $m_z$ = 55 (intermediate depth). \textbf{(b)} Transmission probabilities for very deep 
constrictions, $m_z$ = 8 and different lengths: $L$=10 (upper panel) and $L$ =30 (lower panel).}
\label{fig:Fig3}
\end{figure}

On the other hand,  the edge states, observed here in the energy range from 0.3 eV to 0, drops to $T(E)$ =1 with resonant peaks on top for wide constrictions and anti-ressonances for narrow 
constrictions.  \Fref{fig:Fig3}(a) presents a closer
look at this energy range; for thin armchair steps  ($N_z-m_z = 3$) the red curve shows a group of 10 peaks, these resonances get thinner as the step  increases ($N_z-m_z = 5$).    
Further diminishing $m_z$, the barriers to the 
upper edge contacts become too large, but now the coupling to the lower edge 
becomes relevant. For $m_z \leq 15$ transmission  shows asymmetric Fano-like resonances at the low energy side and sharp 
anti-resonances at higher energies within the central band as will be discussed below. 
It should be recalled here that only  for extremely thin nanoribbons the strong coupling between the edges widens the central band, however, it is not observed in  \Fref{fig:Fig3}(b) a plateau 
enlargement for $m_z=8$ 
because the edges of the  left and right contact are not coupled (keeping the central band of the contacts unaltered).


\begin{center}
\begin{figure} [h]
\includegraphics[width=1\columnwidth]{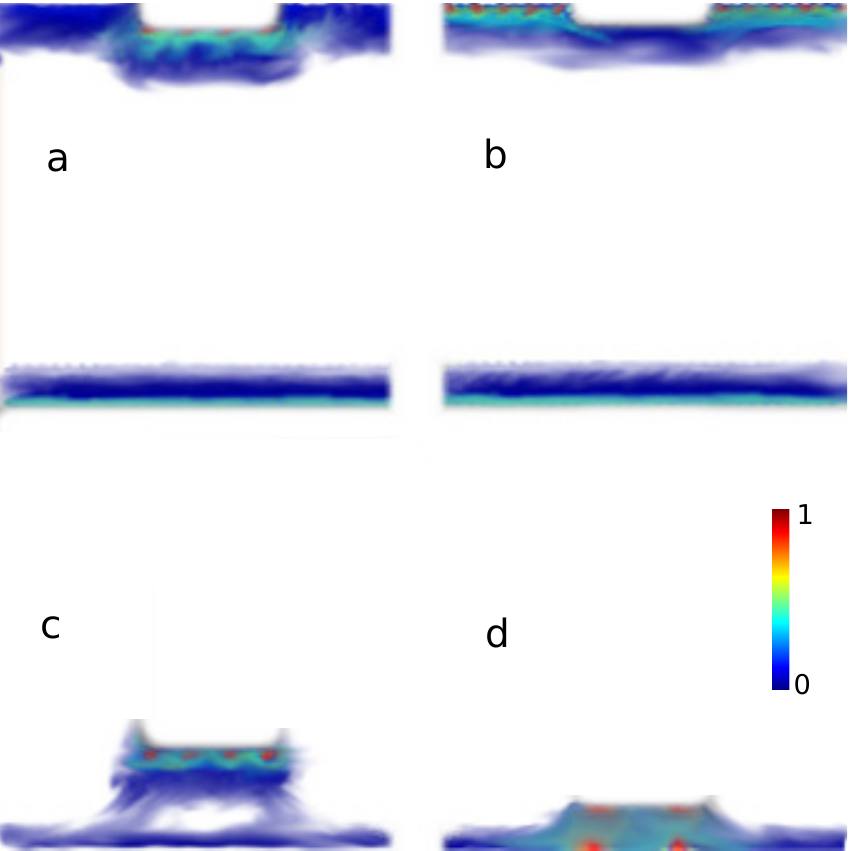}
\caption{(Color online) LDOS at the energies pointed out by arrows, labeled $1$, $2$, $3$ and $4$, in \Fref{fig:Fig3}
for constrictions $L$=10 long. The upper panel is for shallow constrictions, $N_z$ = 60 and $m_z$ = 57: 
\textbf{(a)} at a resonant energy, corresponding to arrow $1$. \textbf{(b)}  Off resonant energy, corresponding to 
arrow $2$. The lower panels are for deep constrictions: \textbf{(c)} at the anti resonance indicated by 
arrow $3$ ($m_z$ = 15); \textbf{(d)} at the anti resonance highlighted by arrow $4$ ($m_z$ =8).}
\label{fig:Fig4}
\end{figure}
\end{center}

In our constriction the role of both channels (discrete states and 
continuum at the upper and lower edges respectively) can be made explicit by 
picturing the local density of states (LDOS) in  \Fref{fig:Fig4}, at the energy values 
indicated by  arrows 1, 	2,  3 and 4 in \Fref{fig:Fig3}. 

In \Fref{fig:Fig4}(a), looking at the LDOS corresponding to a transmission peak pointed out by the arrow 1, it is clear that the higher values of 
LDOS appear on the edges, the 
confined state at the constriction in the upper edge clearly stands out. 
It should be noticed that this LDOS is slightly asymmetric, 
since the structure with $L$ equal to an even number of atoms is asymmetric 
(see  \Fref{fig:Fig2}(a)). This asymmetry leads to a resonance peak $T<
1$ \cite{But88} superposed to the background plateau. 
$L$ equals to an odd number 
of atoms restore complete symmetry, leading to higher resonances, $T \approx 
1$ (not shown here). A less intense LDOS along the lower edge corresponding to 
the $T=1$ plateau can also be observed. The LDOS along the upper edge outside 
the constriction region is less intense in the figure scale, due to 
the prominence of the confined state. Far from a resonance, actually between 
two resonances, situation pointed out by arrow 2 in \Fref{fig:Fig3}(a), the LDOS in the confined part 
of the upper edge is strongly suppressed, enhancing the contribution along the entire 
lower edge and the upper edge contact sections (i.e., outside the confining 
region), \Fref{fig:Fig4}(b). 

A quite different situation is depicted in \Fref{fig:Fig4}(c)-(d), where the confined sate in the constriction is
decoupled from the upper edge with a varible coupling to the lower edge. In \Fref{fig:Fig4}(c), 
corresponding to the anti-resonance labelled as 3 in \Fref{fig:Fig3}, we observe a faint coupling to the lower edge.

\begin{center}
\begin{figure} [h]
\includegraphics[width=0.9\columnwidth]{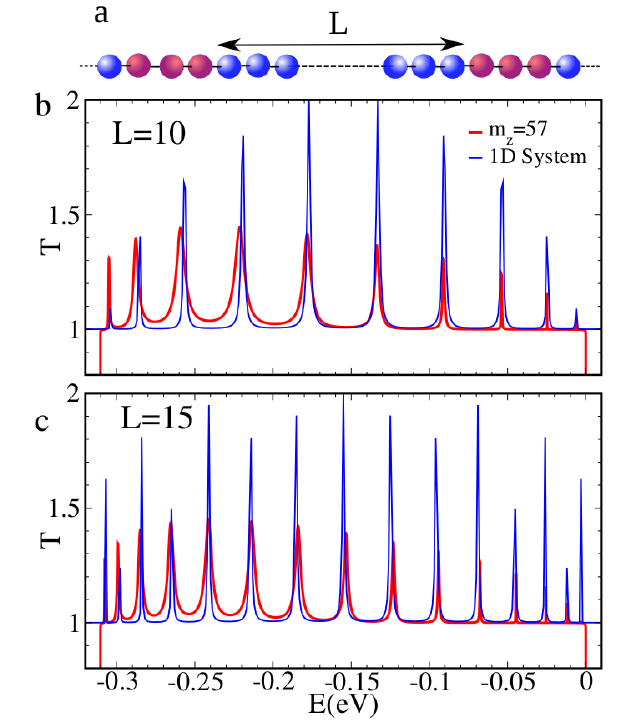}
\caption{(Color online)Comparison between the transmission probabilities at the central band
energy ranges for shallow constrictions, $m_z$ = 57, and  equivalent $s$-like orbitals chain toy 
model. \textbf{(a)} Representation of the one 1D double barrier quantum well structure at the upper edge. the central blue sites 
represent the quantum well at the constriction, while the red ones are 
for the barriers (armchair steps). The left and right contacts at the upper edges are also represented by blue sites.. \textbf{(b)}
Transmission probabilities for a $L$ =10 constriction: 4 band tight binding model (red) and the 
toy model (blue). \textbf{(c)} The same as \textbf{(b)} but for a longer constriction, $L$ =15.}
\label{fig:Fig5}
\end{figure}
\end{center}

The LDOS plots reveal the unique
character of  the resonances observed in \Fref{fig:Fig3}: resonant tunnelling through confined edge states
in the constriction defined by an armchair-like step double barrier structure. 
It should be noticed that the number of resonances is identical to the
number of atoms removed along the segment that define the
length of the constriction, $L$ = 10,
indicating that the transmission shows a spectroscopy of the 1D states
at the edge of the constriction. Indeed, increasing the length of the
constriction will increase at the same proportion the number of the resonances
(not shown here). The fact that the resonances are insensible to the constriction width
(hence the area for a fixed length) is a further indication that we are dealing
with a strictly 1D effect at the edges: albeit the underlying 2D crystal, the
behavior revealed here is the one of an effective atomic chain. Here we should notice that the only signature of the underlying 2D crystal is given by the resonances widths. Recalling 
\Fref{fig:Fig1}(e), the resonances near the bottom of the central band, $Ka \approx 0$, correspond to states which deeper penetrate in the bulk, hence the barriers defined by the device steps are 
less effective than for resonances at higher energies.
Two dimensional structured systems, like nanorribons,  present transmission probabilities with multi channel contributions that are mixed by the geometrical changes along the 
structure\cite{MendSV08,atomtotrasn}. On the other hand, one dimensional systems present single channel transmission probabilities, that are described by s-like orbital chain models. 
In the present 
case,
indeed, the positions of the transmission resonances are shown in \Fref{fig:Fig5} to be well reproduced by a simple
one dimensional double barrier quantum well modelled by a chain of s-like
orbitals \cite{atomtotrasn,ShuG87}. The nearest neighbor hopping ($t_{1D}$) of the one dimensional chain, shown in \Fref{fig:Fig5}(a), is calculated by $|t_{1D}|=\Delta E/4= 0.0775$ eV, where $\Delta 
E$
is the edge states band width obtained from the red curve in \Fref{fig:Fig1}(c). The transmission peaks of the one dimensional quantum well with $L=10$ and $L=15$ atomic sites clearly reproduce the 
position of the resonaces observed for phosphorene constrictions of the same length, see \Fref{fig:Fig5}(b)-(c). The resonant peaks appear at the energies of the infinite square well  energies $E_n = 
2t_{1D} + 2t_{1D}\text{cos}(n\pi/(L+1))$ for $n=1,~2,...,L$

\begin{center}
\begin{figure} [t]
\includegraphics[width=0.7\columnwidth]{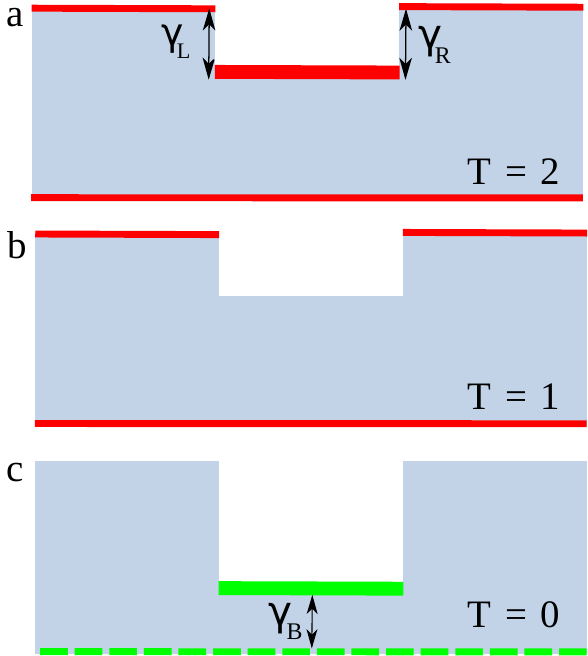}
\caption{(Color online) Schematic representations of the main transmission probability 
conditions. The grey areas represent the nanoribbons with constrictions, with different edge 
states behaviors highlighted by red and green lines. \textbf{(a)} Resonant transmission at the upper 
edge summed up to the continuous transmission at the lower edge, hence $T$ =2. The resonant 
coupling of the confined state at the constriction with the left(right) contact is represented by 
$\gamma_L$($\gamma_R)$. \textbf{(a)} Off resonance negligible transmission at the upper edge with the continuous 
transmission at the lower edge, $T$ =1. \textbf{(c)} Anti-resonance in the transmission due to the strong 
coupling, represented by $\Gamma_B$, between the confined state at the deep constriction and the lower 
edge, now depicted as a dashed line indicating the absence of transmission, $T$ = 0.}
\label{fig:Fig6}
\end{figure}
\end{center}

To recap, the edge states at opposite edge provide one dimensional electronic transport channels embedded in a two dimensional material and the conductance observed can be understood from simple  
model. The resonant peaks on top of the $T=1$ plateau resemble the conductance of two parallel and independent channels,  as shown in \Fref{fig:Fig6}(a). The lower edge provides a 
continuous 
channel of $T=1$ while the upper edge presents a quantum well with tunnelling coefficients $\gamma_{L(R)}$ accross the left(right) barrier defined by the vertical edges 
in 
the figure.  Hence, resonant tunneling becomes only possible when the energy matches the energy of the 
bound states in the well. When this situation is not fulfilled the upper channel is closed and the transmission of the whole system is $T=1$, see  \Fref{fig:Fig6}(b). A deeper step on the 
constriction, on the other hand, would lead to quasi bound states, i.e., wider barriers, connected to the upper edge contacts, but with a significant coupling, $\gamma_B$,  to the 
bottom edge, \Fref{fig:Fig6}(c), 
leading to Fano-like
resonances and antiresonances in the transmission \cite{GhostFano,C5NR03393D}. This situation corresponds to quasi -bound state coupled to a continuum, leadind to fano-like 
asymmetric resonances and antiresonances in the transmission.

The three different line-shapes can be described by a single
expression \cite{HuaLY15}:

\begin{equation}
  \label{eq:TransFano}
  \begin{array}{l}
    T(\epsilon) = \mid t_d\mid^2 \frac{\mid q+\epsilon \mid^2}{1+\epsilon^2}
  \end{array}
\end{equation}

where $t_d$  is the direct transmission without the presence of a scattering region,
$\epsilon =(E-E_R)/\Gamma$, ($E_R$ is the energy of the resonant discrete state
and $\Gamma$ its line width) and $q$ is the Fano asymmetry factor,which represents the ratio of the  resonant tunneling channel to the channel due to the continuum (here represented by the lower edge 
channel). When
$q\rightarrow \infty$ and $t_d\rightarrow 0$ (no continuum channel available) a
a resonance peak develops; for $q \approx \pm 1$ (both channels are relevant), an asymmetric line
shape is revealed, while for $q = 0$ an antiresonance appears.

Recalling the framework of the present work, edge confined states are supported
only by zigzag edges, being absent in armchair or bearded edges \cite{Ezawa,Carvalho}.
Therefore, introducing perturbations to a zigzag edge, like edge vacancies, would locally
destroy these 1D states. The consequences of these perturbations are very relevant
 in the present early stage of phosphorene experimental
development, in which only preliminary steps toward design and realization of
effective devices out of the bulk in the nanoscale have been reported so far.\cite{TayHF2015}.
Hence, effects of the presence of disorder at the edges have to be considered.

\begin{center}
\begin{figure} [h]
\includegraphics[width=0.9\columnwidth]{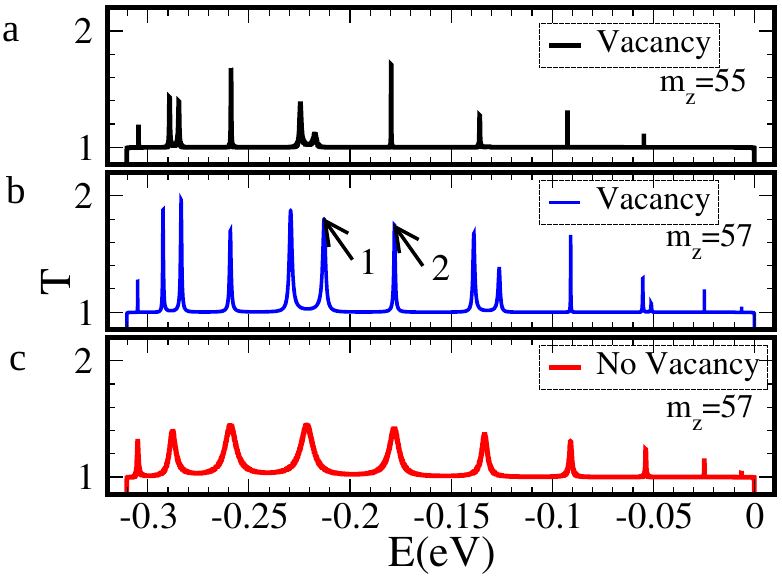}
\includegraphics[width=1\columnwidth]{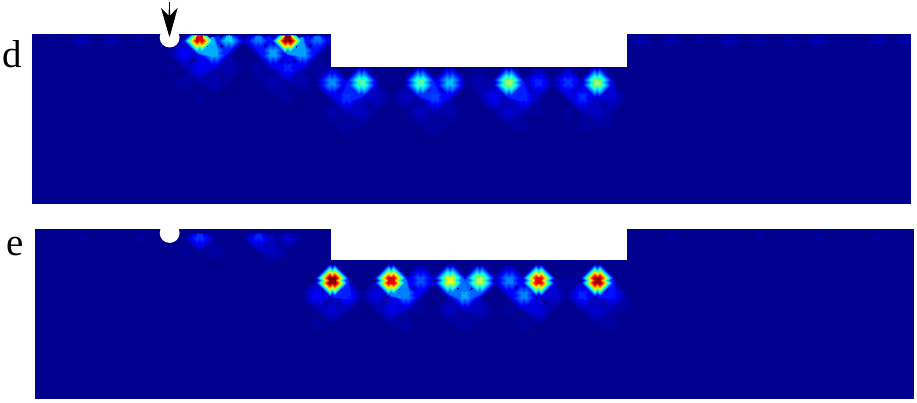}
\caption{(Color online) Edge states transmission probabilities for a nanoribbon $N_z$ = 60 wide, with 
a constriction $L$ =10 long, adding an edge vacancy at the left contact: \textbf{(a)} $m_z$ = 55 and \textbf{(b)} $m_z$ =57. 
\textbf{(c)} The system without the vacancy for the sake of comparison, highlighting the splitting of the 
resonances due to the presence of a vacancy. \textbf{(d)} LDOS at energy indicated by arrow 1 in \textbf{(b)}. 
\textbf{(e)} LDOS at the energy indicated by arrow 2 in \textbf{(b)}.}
\label{fig:Fig7}
\end{figure}
\end{center}

In \Fref{fig:Fig7} we present the transmission probability as a function of
energy as well as LDOS associated to selected resonances in the presence
of vacancies.

Defects are normally seen as mechanisms that hinder the observation of transport
properties associated to shape modulation of nanoscopic low dimensional systems.
Indeed, the resonance spectra are also dramatically modified in the present
case. However, the issue can be seen from an entirely different point of view.
The vacancies change locally the character of the edge, introducing actually
small barriers, further dividing the system into smaller segments. The system chosen here is a device with a vacancy located at the upper edge of the left contact near the central segment (quantum 
well), the exact position is marked by the arrow in \Fref{fig:Fig7}(d). 
What can be
observed in the transmission probabilities in \Fref{fig:Fig7}(a)-(c) is that the resonances of the well at the contact defined to the left by a barrier due to the vacancy, couple to some of the 
states of the original well given by the central segment. Those couplings are identified by the split peaks clearly seen in
\Fref{fig:Fig7}(b). If the barriers defining the central segment(well) are widened, \Fref{fig:Fig7}(a), the splitting diminishes. \Fref{fig:Fig7}(c) depicts the same device in absence of the vacancy 
as a guide for identifying the couplings. The character of the 1D confined states in the presence of a vacancy is illustrated in the LDOS, \Fref{fig:Fig7}(d)-(e), for the resonances corresponding to 
arrows 1 and 2 in \Fref{fig:Fig7}(b), respectively. State 1 corresponds to a double-well (vacancy-upper edge contact-left step-central segment-right step) along the upper edge, exhibiting LDOS at 
both wells, while state 2 is confined mainly to the central quantum well.

Having in mind the previous discussion, while a resonant tunnelling spectroscopy
would become rather involved with the presence of defect induced barriers,
scanning probe microscopy continues a way to reveal the edge quantum
confinement.

\section{Conclusions}

It is inevitable to compare our results with those obtained for graphene constrictions: both zigzag graphene and phosphorene nanoribbons support edge states, however, their signatures on the 
electronic transport properties are completely different. First, edge states in zigzag graphene nanoribbons are sublattice polarized, so one single edge do not contribute to the electron transport 
properties. The 
graphene edge states channel is originated by the overlapping of edge states on opposite edges \cite{1367-2630-11-9-095016}, contrary to what we observe here where a single phosphorene edge provides an independent transport 
channel. Second, localized states in graphene junctions manifest as antiresonances of zero conductance, these states localize over the junction \cite{Wakabayashi:2000cs,Wakabayashi:2001hb,Munoz-Rojas:2006tg}, while the localized states of 
phosphorene 
constrictions strictly  on  grooved zigzag edge and appears as  peaks, asymmetric Fano line shapes or dips in the conductance. In summary, we propose phosphorene zigzag nanorribons as a
platform for
constriction (segment) engineering. In the presence of an engraved segment at the upper edge,
quantum confinement of edge protected states reveals resonant tunnelling transmission 
peaks if the upper edge of the host nanoribbon is uncoupled to the lower
edge. Coupling between edges in thin constrictions give rise to Fano-like and
antiresonances in the transmission spectrum of the system. One could envisage to look these effects by means of transport measurements as well as scanning probe microscopy \cite{PhysRevB.91.115413}. 
The energy scale given by the resonance
spacing is of the order of 10 meV for constriction lengths of $L$ =30 (not shown here),
corresponding to $\sim 5.05$ nm and contacts $\sim 13.3$ nm wide, a benchmark for experimental efforts, recalling that defects may lead to more complex spectra without washing out the main features. 
Other resonant tunnelling mechanisms have been observed in phosphorene nanoribbons with vacancies \cite{jp5110938}, defects \cite{Farooq} and  transverse electric fields \cite{PhysRevB.92.035436}. It 
is important to reinforce that these mechanisms involve states at the interior of the nanoribbon, whereas the effect shown here requires one dimensional states  localized at the edges.

Concomitant to the development of the fascinating physics of 2D
materials, new extreme 1D systems, namely isolated atomic chains, either based
on Carbon \cite{PhysRevLett.102.205501} or metallic elements \cite{Nadj-Perge31102014}, have been obtained and characterized, with their
properties and possible applications theoretically investigated. The present
results suggest a way were effective atomic chains come out from the edges of a
2D crystal.

\section*{Acknowledgments}
CJP and ALCP acknowledge FAPESP, grant
2012/19060-0. DAB acknowledges support from FAPESP grant 2012/50259-8.
PAS acknowledges support from CNPq.
Numerical simulations were performed  at cluster
LaSCADo-UNICAMP, supported by FAPESP under project 2010/50646-6.

\bibliography{bibliograph}

\end{document}